\documentclass[reprint,
prd,onecolumn,amsmath, amssymb, 12pt]{revtex4-2}   
\usepackage[utf8]{inputenc}
\usepackage{mwe,float}
\usepackage{caption,subcaption, url}
\usepackage{natbib}
\usepackage[nointegrals]{wasysym}
\usepackage{rotating}
\usepackage{array,tabu,multirow}
\usepackage{graphicx,bm}
\usepackage{dcolumn}
\usepackage{floatpag}  
\rotfloatpagestyle{empty}
\usepackage{amsmath}
\usepackage{amsthm}
\usepackage{amssymb}
\usepackage{latexsym,upgreek}
\usepackage{hyperref,rotating}
\usepackage{xpatch,bigints}
\usepackage{booktabs,subcaption}
\usepackage{ulem}
\usepackage{textgreek}
\usepackage{mathrsfs}
\makeatletter
\AtBeginDocument{\xpatchcmd{\@thm}{\thm@headpunct{.}}{\thm@headpunct{\enskip}}{}{}}
\makeatother 
\usepackage{gensymb}
\usepackage{mathdots}

\usepackage[version=4]{mhchem}

\usepackage{graphicx,kecpm,slashed}
\usepackage{makeidx}
\makeindex
\usepackage{multirow}
\usepackage[dvipsnames]{xcolor}

\newcommand{\grad}{\nabla}

\newcommand{\bex}{\begin{example}}
\newcommand{\eex}{\end{example}}
\newcommand{\besp}{\begin{split}}
\newcommand{\ensp}{\end{split}}

\newcommand{\om}{\omega}
\newcommand{\f}{\phi}

\newcommand{\by}{\times}

\newcommand{\bos}{\boldsymbol}
\newcommand{\tbf}{\textbf}

\newcommand{\btab}{\begin{tabular}}
\newcommand{\etab}{\end{tabular}}
\newcommand{\barr}{\begin{array}}
\newcommand{\earr}{\end{array}}
\newcommand{\bpm}{\begin{pmatrix}}
\newcommand{\epm}{\end{pmatrix}}
\newcommand{\bit}{\begin{itemize}}
\newcommand{\eit}{\end{itemize}}
\newcommand{\ben}{\begin{enumerate}}
\newcommand{\een}{\end{enumerate}}
\newcommand{\bct}{\begin{center}}
\newcommand{\ect}{\end{center}}

\newcommand{\ra}{\rangle}
\newcommand{\la}{\langle}
\newcommand{\bes}{\begin{split}}
\newcommand{\ens}{\end{split}}

\newcommand{\lt}{\left}
\newcommand{\rt}{\right}

\newcommand{\edoc}{\end{document}}

\begin{document}
\title{Spacetime uncertainty makes quantum field theory finite}

\author{Kevin Cahill}
\email{cahill@unm.edu}

\affiliation{Department of Physics and Astronomy\\
University of New Mexico\\
Albuquerque, New Mexico 87106}
\date{\today}

\nopagebreak

\begin{abstract}

Since Einstein's equations $G_{ij} = 8\pi \, G \,  T_{ij} \, / c^4 $ relate the metric $g_{ij}$ of spacetime to the energy-momentum tensor $T_{ij}$ which is a quantum field, the metric $g_{ij}$ must be a quantum field.  And since the metric $g_{ij}(x)$ is the dot product $g_{ij}(x) = \partial_i p^\alpha(x)  \, \partial_j p_\alpha(x)$ of the derivatives of the points $p(x)$ of spacetime, \textbf{spacetime must be a quantum field}.  Its points have average values $\langle p(x) \rangle$ that obey general relativity and fluctuations $q(x) = p(x) - \langle p(x) \rangle$ that obey quantum mechanics.  It is suggested that the fields of quantum field theory be regarded not as functions $\phi(x)$ of their classical coordinates $x$ but as functions $\phi(p(x))$ of their quantum coordinates $p(x)$. In empty flat spacetime where $p(x) = x + q(x)$ and $x = (t, \boldsymbol x)$, the Fourier exponentials $\exp(i k(x+q(x))$ averaged over normally distributed fluctuations $q(x)$ are gaussians $\exp(i kx -\ell^2 \boldsymbol k^2 - \ell^2 m^2/2)$.  These gaussians make Feynman diagrams finite.   The zero-point energy density of the vacuum also is finite --- but negative and too large to explain dark energy unless new bosons exist.

\end{abstract}
 
\nopagebreak

\maketitle

\nopagebreak

\section{General relativity made 
space physical}
\label{General relativity made 
space physical}

\par
Einstein's equations 
$G_{ij} = 
8\pi \, G \,  T_{ij} \, / c^4 $
relate his tensor 
$\, G_{ij} \equiv {}  
R_{ij} - \thalf g_{ij} R \,$
which is a sum of products of the 
metric $g_{ij}$, 
its inverse, and its derivatives 
to the  
energy-momentum tensor 
$T_{ij}$ which is a sum of products 
of quantum fields and their derivatives.
Thus the metric $g_{ij}$
must be a quantum field. 
\par
Our curved  spacetime
can be embedded 
isometrically in a flat 
higher-dimensional semi-euclidian
spacetime 
whose points
$p^\a(x)$ have one dimension
of time and at most 19 dimensions of 
space~\citep{Nash1956, MR0262980, Clarke1970, Guenther1989, MuellerSanchez2011, Ake:2018dzz, CahillSec13.12}.
Their derivatives $ \p_i p^\a(x)$ 
with respect
to the 4 coordinates $x^i$ 
of spacetime 
are 4 vectors 
\begin{equation}
v^\a_i = \frac{\p p^\a(x)}{\p x^i} 
= \p_i p^\a(x)
\end{equation}
in which $\a$ runs from 0 to $ s \le 19$.
The 16 dot products of 
these four vectors
are the 16 components of the 
metric
$g_{ij}(x) $
\begin{align}
g_{ij}(x) = {}&
\p_i p^\a(x)
\,
\p_j p_\a(x)
=  {}
-
\p_i p^0(x)
\,
\p_j p^0(x)
+
\sum_{k=1}^s
\p_i  p^k(x) 
\,
\p_j p^k(x) .
\label {metric is dot product}
\end{align}
Since the metric $g_{ij}(x)$ 
is a quantum field 
made of 
dot products 
of derivatives
of points of spacetime,
\tbf{spacetime itself
must be a quantum field}.

\par  
Each point $p(x)$ of spacetime 
has an average value
$\la p(x) \ra $ 
given in terms of the 
appropriate density operator $\rho$
as the trace of 
$\la p(x) \ra = 
\tr \lt( p(x) \, \rho \rt) $.
The average value $\la p(x) \ra$
obeys general relativity.
The remainder 
$q(x) = p(x) - \la p(x) \ra$
obeys quantum mechanics.
It is suggested that 
the fields 
of quantum field theory
be regarded not as functions $\f(x)$ 
of their classical coordinates $x$
but as functions $\f(p(x))$ of 
their quantum coordinates $p(x)$. 

\par

In empty flat spacetime,
the points are 
$p(x) = x + q(x)$ 
with $x=(t, \bos x)$, 
and the fields $\f(p(x))$ 
are then  integrals 
of the Fourier exponentials
$\exp(i k p(x)) = \exp(ik (x + q(x)))$.
These exponentials 
when averaged over 
normally distributed 
quantum fluctuations $q(x)$ 
of standard deviation $\ell$ 
are gaussians 
$\exp(i k x - \ell^2 \boldsymbol k^2 - \tiny{\frac{1}{2}} \ell^2 m^2)$
that make Feynman diagrams finite. 
The zero-point energy density
of the vacuum also 
is finite, but negative
and much greater than 
the observed 
dark-energy density  
($3 \by 10^{-47}\,\text{GeV}^4$)
unless new bosons exist. 

\section{Points in empty flat  spacetime}
\label{Points in empty flat spacetime}

\par
The implementation 
of these ideas is simplest
in empty flat  spacetime
where a point
$p(x)$ 
may be taken to be the sum of its
average value 
$\la p(x) \ra = x \equiv (t, \boldsymbol x)$
and its quantum-mechanical
part $q(x)$ 
\begin{equation}
p(x) = \la p(x) \ra + q(x) = x + q(x) .
\label{a point is}
\end{equation}
Each quantum-mechanical
part $q^i(x)$ 
will be assumed to be distributed 
according to its 
wave function in the vacuum
$\la q^i(x) | 0 \ra$.
Since the quantum behavior
of spacetime has not yet been observed,
the probability
distribution $|\la q^i(x) | 0 \ra|^2$ 
must be small for 
$q^{i2}(x) > \ell^2$
in which $\ell$ is a short 
distance less than, let us say,
$10^{-21}\,\text{m}$ (1\,zm). 
\par
The conclusions of this paper
are largely independent of the 
detailed form of the 
distribution $|\la q^i(x) | 0 \ra|^2$
as long as it is 
smooth, symmetric 
about $q^i(x) = 0$,
independent of $x$,
and integrable with
\begin{equation}
1 = \int_{-\infty}^\infty dq^i \,\, |\la q^i(x) | 0 \ra|^2  .
\label{square-integrable}
\end{equation} 
The simplest choice is also 
independent of $i$ 
\begin{align}
| \la q^i(x) |  0  \ra |^2
= {}&
\frac{1}{ \sqrt{2\pi} \, \ell}
\exp\lt( 
- \frac{q^{i2}(x)}
{2 \ell ^2}
\rt) 
\label {a 4 gaussian distribution}
\end{align}
and has a standard deviation $\ell$ 
is less than 
1\,zm $ \approx 
\hbar c /(200 \, \text{TeV})$.
The probability distribution 
for $q(x) =(q^0, \bos q(x))$ is
\begin{equation}
|\la q(x) | 0 \ra|^2 ={}
|\la q^0(x) | 0\ra \la q^1(x) | 0\ra
\la q^2(x) | 0\ra \la q^3(x) | 0\ra|^2.
\end{equation}  
This probability distribution 
violates special relativity 
on scales shorter than 1 zm.
No Lorentz-invariant function of $q$ is
square integrable.

\par

The average values of the 
even monomials $q^i(x)^{2n}$ are 
\begin{equation}
\begin{split}
\la q^i(x)^{2n} \ra _\ell 
={}&
\int_{-\infty}^\infty 
q^i(x)^{2n} 
| \la q^i(x) |  0  \ra |^2 
\, dq^i(x) 
={}
(2n-1)!! \,\, \ell^{2n}
\end{split}
\end{equation} 
while those 
of the odd monomials vanish,
$\la q^i(x)^{2n+1} \ra_\ell
= 0$.
These average values are
independent of $x$ because
although each point $x$ 
has its own probability distribution
$\la q(x)|0\ra$, they are all the same
in empty flat spacetime.
So their spacetime derivatives vanish,
$\p_i \la q^j(x)^n \ra = 0$. 
The parameter $x$ in $q^i(x)$
labels the $q$ associated with 
the point $x$; it is not an argument.

\section{Fields in empty flat  spacetime}
\label{Fields in flat empty spacetime}

\par
In empty flat spacetime with  
$g_{ij}(x) = \eta_{ij}$,
the exponential 
$\exp( \pm i  k \cdot  (x + q(x) ))$
averaged over the quantum
fluctuations 
$q(x) = (q^0(x), \boldsymbol q(x))$
is
\begin{align}
\big \la 
\exp( \pm i \, k \cdot 
( {x + q(x)} ))
\big\ra_\ell
={}&
\int d^4   q(x) \,\, 
| \la q(x) | 0 \ra |^2 \,
\exp( \pm i  \,  
k \cdot   ({x + q(x)}) )
\nn\\
={}&
\int \frac{d^4 q(x) }
{ (2\pi)^{2} \ell^4}
\,
\exp\big( \pm i  k \cdot  ( {x + q(x)} ) 
- q^2_e(x)/2 \ell^2\big)
\label{averaged exponential}\\
={}&
\exp\big( \pm i {k \cdot  x}
- \thalf \ell^2  k_e^2 \big) 
\nn
\end{align} 
in which $k_e^2 = {} 
k^{02} + \boldsymbol k^2$. 
The averaged exponential 
$\big \la 
\exp( \pm i \, k \cdot 
({x + q(x)} ))
\big\ra_\ell$
is the original exponential 
$\exp( \pm i \, k \cdot x )$
damped by 
$ \exp( - \ell^2 k_e^2/2) $.
\par
The usual Fourier decomposition
of a spin-zero field of mass $m$
in flat empty space is
\begin{align}
\f(x) = {}& \int 
\frac{d^3k}{\sqrt{(2\pi)^3 2 \om_k}} 
\, \Big[ 
e^{i k  x} a(k) 
+ e^{-i k x} 
a^\dag(k) \Big]
\label {the usual field}
\end{align}
in which
$k  x  =
\boldsymbol k \cdot \boldsymbol x - \om_k x^0$
and $\om_k = \sqrt{\boldsymbol k^2 + m^2}$. 
It follows that 
the field averaged over
the quantum fluctuations is 
\begin{align}
\big \la 
\f ( x )
\big \ra_{\ell} 
= {}& \int 
\frac{d^3k \, d^4q(x) \, \, 
| \la q(x) | 0 \ra |^2}
{\sqrt{(2\pi)^7 
\, 2 \om_k}} 
\, \Big[ 
 e^{i k (x + q)} \, a(k)  
+ e^{-i k (x + q)} \, 
a^\dag(k)  \, \Big]  
\nn\\
= {}& 
\int 
\frac{d^3k}{\sqrt{(2\pi)^3 2 \om_k}} 
\, 
\Big[ 
e^{i k x} a(k) 
+ e^{-i k x } 
a^\dag(k) \Big]  
\, e^{- \ell^2 k_e^2 /2} 
\label {the 4 ell field}
\end{align}
in which the $k$ integral
is damped by
$\exp(- \ell^2 k_e^2 /2) = 
\exp(- \ell^2(\boldsymbol k^2 + \thalf m^2))$.
The averaged field 
$ \la \f ( x ) \ra_{\ell}$ 
is the usual field (\ref{the usual field})
but with each Fourier component
damped by 
the exponential
$\exp( - \ell^2 k_e^2/2 )$
where
$k_e^2 = m^2 + 2 \boldsymbol k^2$.
In the limit $ \ell \to 0$,
the averaged field 
$ \la \f ( x ) \ra_{\ell}$ 
is the
usual field
(\ref{the usual field}).

\par

The average value of
a field $\psi_{ab}$ of any spin  
also is damped at high momenta
by the same exponential
\begin{equation}
\big \la 
\psi_{ab} ( x )
\big \ra_{\ell} 
= 
\sum_\s \int 
\frac{d^3k}{\sqrt{(2\pi)^3 2 \om_k}} 
\, 
\Big[ 
u_{ab}(p,\s)
e^{i k x} a(k) 
+ 
v_{ab}(p,\s)
e^{-i k x } 
a^\dag(k) \Big]  
\, e^{- \ell^2 k_e^2 /2} .
\end{equation}  

\par

In what follows,
the abbreviation
$
\f_{\ell}( x ) \equiv
\la \f ( x ) \ra_\ell
$ will be used.

\section{Metric in empty flat spacetime}
\label{Metric in empty flat spacetime}

\par

The metric $g_{ij}(x)$ depends 
(\ref{metric is dot product})
upon the derivatives
$\p_i p^\a(x)$ of the points
$p(x) = \la p(x) \ra_\ell + q(x)$
and therefore upon
the derivatives
of the fluctuations $q(x)$.
But in empty flat spacetime 
$ \la p(x) \ra_\ell = {} (t, \bos x) $,
$ \p_i  \la p^j(x) \ra_\ell = \d^j_i $,
and
$ \p_i  \la q^j(x)^n \ra_\ell = 0 $,
so I will take the metric to be 
$\eta_{ij}$ and set 
$k^i g_{ij}(x+q) (x^j + q^j)
=
k^i \eta_{ij}  (x^j + q^j)$.

\section{Examples}
\label{Examples}

\subsection{Products of fields in flat space}
\label{Products of fields in flat space}

A quantum theory
of a single spin-zero field $\f(x)$
is defined by
the average values 
in the vacuum 
of products of the field
$\la 0 | \f(x_1) \, \f(x_2) \dots
\f(x_n) | 0 \ra$.
If we replace each field 
$\f$ by $\f_\ell$,
then we get 
\begin{align}
\big \la \la 0 | \f(x_1) \, \f(x_2) \dots
\f(x_n) | 0 \ra \big \ra_\ell
& = {} 
\la 0 | \int 
\frac{d^3k_1}
{\sqrt{(2\pi)^3 2 \om_{k_1}}} 
\, e^{- \thalf \ell^2  k_1^2} \,
\Big[ 
e^{i k_1 x_1} a(k_1) 
+ e^{-i k_1 x_1 } 
a^\dag(k_1) \Big]  
\nn\\
\by {}&
\int 
\frac{d^3k_2}
{\sqrt{(2\pi)^3 2 \om_{k_2}}} 
\, \, e^{- \thalf \ell^2  k_2^2} \,
\Big[ 
e^{i k_2 x_2} a(k_2) 
+ e^{-i k_2 x_2 } 
a^\dag(k_2) \Big]  
\label{n averaged fields}
\\
{}& \dots \, \by 
\int 
\frac{d^3k_n}
{\sqrt{(2\pi)^3 2 \om_{k_n}}}
\, \, e^{- \thalf \ell^2  k_n^2} \,
\Big[ 
e^{i k_n x_n} a(k_n) 
+ e^{-i k_n x_n } 
a^\dag(k_n) 
\Big]  
|0\ra   .
\nn
\end{align}

\par
All such damped functions
are finite functions 
of $ x_1, x_2, \dots, x_n$.

\subsection{The fundamental equal-time  commutation relation}
\label{The fundamental equal-time  commutation relation}
\par
The equal-time  commutator
of the  field
$\f_\ell ( x )$ 
as given by (\ref{the 4 ell field}) 
with its conjugate momentum 
$\pi_\ell = 
\dot \f_\ell ( x )$ 
averaged over
the $q$'s is
\begin{equation}
\begin{split}
\boldsymbol [ \f_{\ell}(t, \boldsymbol x ) , 
\, \pi_{\ell}(t, \boldsymbol y ) \bos]
={}&
\boldsymbol {\bigg[}
\frac{d^3k}{\sqrt{(2\pi)^3 2 \om_k}} 
\, e^{- \thalf \ell^2  k_e^{2}}
\Big( 
e^{i k \cdot x} a(k) 
+ e^{-i k \cdot x } 
a^\dag(k) \Big),
\\
{}&
\frac{d^3k'}{\sqrt{(2\pi)^3 2 \om_{k'}}} 
\,  i \om_{k'} e^{- \thalf \ell^2  k_e'^2}
\Big( 
- e^{i k' \cdot y} a(k') 
+ e^{-i k' \cdot y } 
a^\dag(k') \Big) 
\boldsymbol {\bigg]}
\label{commutator}\\
={}&
i\int \frac{d^3k}
{(2\pi)^3}
e^{ 
i \boldsymbol k \cdot (\boldsymbol x - \boldsymbol y ) 
- \thalf \ell^2  k_e^2} 
={}
\frac{1}{(2\ell\sqrt{\pi})^3}
\exp\Big[
- \frac{(\bos x - \bos y)^2}{4 \ell^2}
- \frac{\ell^2 m^2}{2}
\Big]
\end{split}
\end{equation} 
which approaches the usual result
$ i \, \d(\boldsymbol {x-y}) $
as $\ell \to 0$ 
but is finite for all
$\boldsymbol x$ and $ \boldsymbol y$.

\subsection{Momentum conservation and
cross-sections damped at high energy} 
\label{Conservation of momentum} 

Conservation of momentum
is due to the homogeneity 
of spacetime. 
The lowest-order 
$2 \to 2$ amplitude
in $\l \, \f_{\ell}^4$ theory is
\begin{equation}
\begin{split}
A( p, k \to p', k' )_{\ell}
={}&
- i \frac{\l}{4!}
\int  d^4x \, \,
\la p', k' |
\f_{\ell}^4 (x)
| p, k \ra
\\
={}&
- i \l \,\, \frac{\d(k + p - k' - p')}
{(2\pi)^2 \, 4 \, 
\om_p \om_k \om_p' \om_k'} \, \,
e^{- \ell^2(k_e^2 + p_e^2
+ k_e'^2 + p_e'^2)/2}
\label{2to2 amplitude}
\end{split}
\end{equation} 
which conserves 4-momentum.
This and other amplitudes 
go to zero 
at high momentum and 
may let us measure the 
parameter $\ell$ if it is closer 
to one zeptometer (1 zm $\simeq 1/ 200$ TeV) than to the Planck length.

\subsection{Feynman's propagator}
\label{Feynman's propagator}

The Delta function
$ \Delta_{+,\ell}(x-y) $
is the commutator 
\begin{align}
\Delta_{+,\ell}(x-y) 
={}&
\boldsymbol [ \f_{\ell}^+(x), \f_{\ell}^-(y) \boldsymbol ]
=
\int \frac{d^3k}{(2\pi)^3 2 \om_k} \, 
e^{i \boldsymbol k \cdot (\bos{x-y}) 
- i \om_k (x^0-y^0) } \,
e^{- \ell^2 k_e^2} 
\label {Delta = [,]}
\end{align} 
and Feynman's propagator 
$ \Delta_{F, \ell}(x-y) $
is then 
\begin{align}
\Delta_{F, \ell}&(x-y) ={}
i \langle 0 | 
\mathcal{T}\!\left \{ 
\phi_{\ell}(x) \phi_{\ell}(y) \right\} 
| 0 \ra
={}
\int \frac{d^4k}
{(2\pi)^4 }
\frac{e^{i k ( {x-y} ) 
- \ell^2 k_e^2}}
{k^2 + m^2} 
\nn
\end{align}
which is the usual propagator
damped at high momentum by
$\exp(- \ell^2 k_e^2)$. 

\subsection{Loops}
\label{Loops}

In $\l \f^4$ theory,
the  self-energy diagram 
of lowest order 
is a linearly divergent integral 
multiplied by an energy-momentum
conserving delta function. 
In $\l \f^4_\ell$ theory, 
it is proportional to 
the convergent integral
\begin{equation}
\begin{split}
L ={}& 
-i \l \d(p' - p)  
\int 
\frac{d^3k }
{(2\pi)^2 4 \om_{k}\om_{p}} 
\, 
e^{-  2\ell^2 (\bos{p^2 + k^2  + m^2)}} .
\end{split} 
\end{equation} 
multiplied by the 
delta function
$\d(p' - p)$.

\par
Other loop integrals also 
converge because of the
exponentials $\exp(- \ell^2 \bos k^2)$.

\subsection{Quantum Gravity}
\label{Quantum Gravity}

Perturbative general relativity 
is nonrenormalizable 
because it
involves infinitely many 
differently divergent 
Feynman diagrams
and so requires
requires infinitely many
different counterterms,
but all such diagrams
are finite 
if the metric
$g_{ij}(x)$ is replaced by
\begin{equation}
g_{ij \, \ell}(x) ={}
\int 
\frac{d^3k}{\sqrt{(2\pi)^3 2 \om_k}} 
\, 
\Big[ 
e^{i k x} a_{ij}(k) 
+ e^{-i k x } 
a_{ij}^\dag(k) \Big]  
\, e^{- \ell^2 k_e^2 /2} 
\end{equation}
which is damped exponentially
at high momentum.

\subsection{Dark energy}
\label{Dark energy}

The hamiltonian of a
single free spin-zero field 
of mass $m$ is
\par
\begin{align}
H = {}&
\half  \int  d^3x \,
\Big( \pi^2(x) + (\grad \f(x))^2 
+ m^2 \f^2(x)  \Big)
\end{align}
in which $\pi =\dot \f$
and all fields are at $t=0$.
Its mean value
in the flat-space vacuum 
 has an energy density
\begin{align} 
\rho_m ={}&
\half \int \frac{d^3k}
{(2\pi)^3} \,\, 
\sqrt{\boldsymbol k^2 + m^2}
\end{align}
that is quartically divergent.

\par

The zero-point energy of 
the field
$\f_{\ell}(x)$, however, 
is 
\begin{align}
\la 0 | H |  0  \ra
={}&
\half  \int d^3x \, 
\la 0  | \dot \f_{\ell}^2(x) + 
(\grad \f_{\ell}(x))^2 
+ m^2 \f_{\ell}^2(x)  | 0  \ra
 \nn\\
={}&
\half  \int \frac{d^3x \, d^3k}
{(2\pi)^3 \, 2\om_k}
\Big( 
\om_k^2 + k_e^2
+ m^2
\Big) \, e^{- \ell^2  k^{i2}x^{k2}} 
=
\int \frac{d^3x \, d^3k}
{2(2\pi)^3 } \,
\om_k \, e^{- \ell^2  k^{i2}x^{k2}} .
\end{align} 
Its energy density 
is finite and proportional
to a modified Bessel function
\begin{align}
\rho_{m,\,\ell}
={}&
\int  \frac{d^3k}
{2(2\pi)^3 } \, 
\sqrt{m^2 + \boldsymbol k^2} \,\,
 \,\,
e^{- 2\ell^2 \boldsymbol k^2 - m^2 \ell^2}
={}
\frac{m^2 K_1(\ell^2 m^2)}
{32 \pi^2 \ell^2} .
\label {Energy density of a single 4 epsilon mode}
\end{align}

\begin{figure}[h!]
\begin{center}
\includegraphics[width=5.2in, trim={0.5in 2.6in 0in 2.5in}, clip]
{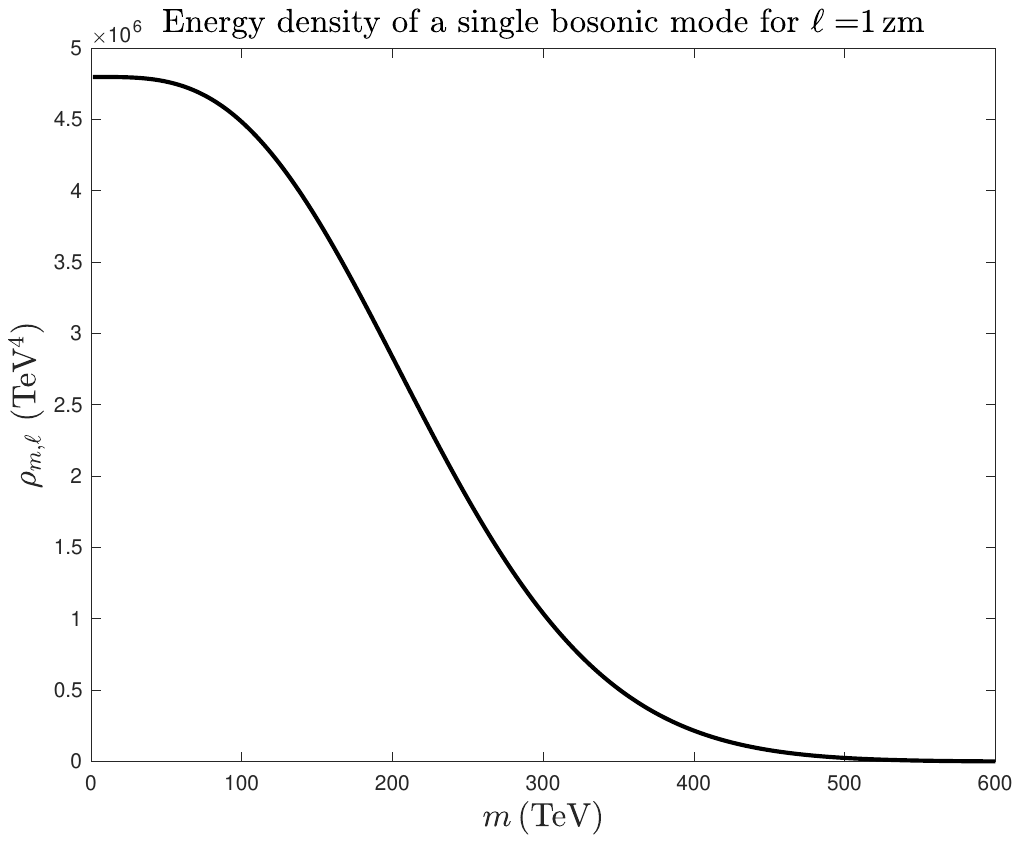}
\caption{The energy density $\rho_{m, \ell}$
of a single bosonic mode 
(\ref{Energy density of a single 4 epsilon mode})
is plotted
for $\ell = 1$\,zm and 
$m \leq 600$\,TeV.} 
\label {besselk}
\end{center}
\end{figure}

\par
The energy density
of every mode $i$ 
of a boson of mass $m$
is finite and given by 
(\ref{Energy density of a single 4 epsilon mode})
which is
$\rho_{0,\,\ell}
={}
1/(32 \pi^2 \ell^4)$
for $m=0$. 
If the field is fermionic, then 
the energy density of each mode
is $- \rho_{m,\ell}$.
The energy density 
$\rho_{m, \ell}$
(\ref{Energy density of a single 4 epsilon mode}) 
is plotted in the figure 
(\ref{besselk}) 
for  $\ell = 1$\,zm
and $m < 600$ TeV.

\par

The contribution of the
zero-point energy density
of the vacuum to
the density of dark energy 
is therefore the sum of $\rho_{m, \ell}$
over all boson modes $b$
minus the sum of $\rho_{m, \ell}$
over all fermion modes $f$
\begin{align}
\rho_{\ell}
={}&
\frac{1}{32 \, \pi^2 \ell^2}
\lt( \sum_{b=1}^B
m_b^2 \, K_1\big(\ell^2 m_b^2 \big)
-
\sum_{f=1}^F
m_f^2 \, K_1\big(\ell^2 m^2_f  \big)
\rt) .
\label {dark density theory 4}
\end{align}
This formula 
(\ref{dark density theory 4})
requires knowledge of
the unknown distance $\ell$
and of 
the masses and spins
of all particles including
those that are dark 
or inaccessibly heavy. 
So its predictive value 
is limited.

\par

But we can tentatively draw 
certain conclusions.
If all the particles are massless
on a scale of 200 TeV
and $\ell \leq 1$\,zm,
then the energy density
(\ref{dark density theory 4})
is approximately 
$
\rho_{\ell, 0}
={}
(B - F)/(32 \, \pi^2 \ell^4)$. 
There are 96 known fermion modes
and only 30 known boson modes,
so $\rho_{\ell, 0}$ 
is about
$- 4.7 \by 10^{18} \, \text{GeV}^4$
unless $B=F$.
Thus if the energy density
$\rho_{\ell, 0}$
is to approximate the observed 
density of dark energy,
$ 3 \by 10^{-47}\,\text{GeV}^4$,
then there must be at least 
66 new undiscovered boson modes 
(e.g., 22 massive vector bosons)
so that $B=F$.
They presumably  
are dark or too heavy
to have been seen.  
\par
If $B = F$, then the leading term 
in the energy density 
$ \rho_{\ell} $ 
is proportional
to the alternating quartic sum 
\begin{equation}
\frac{1}{128\pi^2}
\sum_i (-1)^{2j_i}  m_i^4
\big(
2 \c - 1 
+ 2 \log \big(\thalf \ell^2 m_i^2\big)
\big)
\label{vanishing leading term}
\end{equation} 
which is much too large 
unless there are remarkable cancellations.
But if this sum (\ref{vanishing leading term}) 
and $B-F$ should vanish, then
the resulting energy density 
\begin{equation}
\rho_{\ell} ={}
\frac{\ell^4}{2048 \pi^2}
\sum_i (-1)^{2j_i}  
m_i^8
\Big[
4\c - 5 +4 \log\big(\thalf \ell^2 m^2\big)
\Big] 
\equiv  \frac{\ell^4}{2048 \pi^2} \,
M^8  . 
\end{equation} 
would approximate
the observed density of dark energy 
$3 \by 10^{-47}\,\text{GeV}^4$
if the length $\ell$ were 
\begin{equation}
\begin{split}
\ell ={}& 
\frac{5.5 \by 10^{-27} \,\text{m}}{\lt(M/\text{GeV}\rt)^2}
=
\frac{3.4 \by 10^8}
{\lt(M/\text{GeV}\rt)^2}
\,\, \ell_P.
\label {the condition}
\end{split}
\end{equation}
So agreement with observations
requires 
the vanishing of both $B-F$ and of 
the quartic alternating sum 
(\ref{vanishing leading term}) 
as well as a value of the length $\ell$ 
equal to $\ell \approx 3.4 \by 10^8 \, \ell_P
/(M/\text{GeV})^2$.

\section{Summary}
\label{Summary}

Since Einstein's equations $G_{ij} = 
8\pi \, G \,  T_{ij} $ 
relate the metric $g_{ij}$
to the energy-momentum tensor
$T_{ij} $ which is a sum 
of products of
quantum fields,
the metric must be a quantum field.
And since the metric 
is the dot product 
(\ref{metric is dot product})
of derivatives of 
the points $ p(x) $ of 
spacetime 
$g_{ij}(x) = \p_i p^\a(x)  \, \p_j p_\a(x)$,
spacetime must be a quantum field.
In empty flat spacetime, a point $p(x)$
is the sum
$p(x) = \la p(x) \ra
+ q(x)$
of its average value 
$\la p(x) \ra = x$ 
and its quantum part $q(x)$
which obeys quantum mechanics.
\par
It is suggested that 
the fields 
of quantum field theory
be regarded not as functions $\f(x)$ 
of their classical coordinates $x$
but as functions $\f(p(x))$ of 
their quantum coordinates $p(x)$ 
which in empty flat spacetime
are $p(x)=x+q(x)$. 
The fields $\f(p(x))$ 
are then  integrals 
of the Fourier exponentials
$\exp(i k p(x)) = \exp(ik (x + q(x)))$.
If the probability distribution 
$| \la q(x) | 0 \ra |^2$
of the quantum part 
$q(x)$ is normal and 
of standard deviation $\ell$, then
the Fourier exponential
$\la \exp(i k ( x + q(x) ) \ra$
integrated over $q(x)$
is the gaussian 
$ \exp( i k x - \ell^2  \boldsymbol k^2 - \thalf \ell^2 m^2) $.
These gaussians make 
Feynman diagrams finite,
including those of general relativity.
They also shrink cross-sections
at high momentum, which 
may provide
a way to test the proposed physics
and to measure the length $\ell$.
The zero-point energy density
of the vacuum also is finite,
but huge and negative unless 
there are 66 new boson modes. 
To get agreement with the
observed value $3 \by 10^{-47}$ 
of the density of dark energy,
the alternating quartic sum 
(\ref{vanishing leading term})
must also vanish, 
and the parameter $\ell$
must satisfy the condition
(\ref{the condition}).
The physics proposed here
should be compared
with conventional renormalization
and applied to the analysis
of high-precision experiments
such as measurements 
of electric dipole moments.

\section{Acknowledgments}
I am grateful to Rouzbeh Allahverdi, 
Michael Grady,
David Kaiser, Malcolm Perry, and
Barmak Shams Es Haghi for helpful conversations
and email.

\bibliography{physics,math,lattice,books}
 
\end{document}